\documentclass[11pt]{article}

\usepackage[T1]{fontenc}
\usepackage{lmodern}
\usepackage[utf8]{inputenc}
\usepackage[margin=1in]{geometry}
\usepackage{graphicx}
\usepackage{booktabs}
\usepackage{xcolor}
\usepackage{authblk}
\usepackage[numbers,sort&compress]{natbib}
\usepackage{url}
\usepackage{listings}
\usepackage{amsmath}
\usepackage{amssymb}
\usepackage{amsthm}
\usepackage{multirow}
\usepackage{array}
\usepackage{microtype}
\usepackage{tikz}
\usetikzlibrary{positioning,arrows.meta,fit,calc,backgrounds}
\usepackage[section]{placeins}
\usepackage{float}
\usepackage{hyperref}

\definecolor{darkblue}{rgb}{0.0,0.0,0.45}
\definecolor{accent}{HTML}{9A8FC0}
\definecolor{greyblue}{HTML}{BCCADB}
\hypersetup{colorlinks=true, citecolor=darkblue, linkcolor=darkblue, urlcolor=darkblue, pdftitle={VulnBump: A Benchmark for Fixing Vulnerable Dependencies in Real Projects While Keeping the Build and Tests Green}}

\theoremstyle{definition}

\newcommand{\bench}{VulnBump}
\newcommand{\remy}{Evo Remediation Agent}
\newcommand{\aided}{Claude Code + Snyk scan}
\newcommand{\unaided}{Claude Code unaided}
\newcommand{\pmbump}{package manager bump}
\newcommand{\Pmbump}{Package manager bump}

\title{\bench{}: A Benchmark for Fixing Vulnerable Dependencies in Real Projects While Keeping the Build and Tests Green}

\author[1]{Gabija Pasiunaite}
\author[1]{Ryan McMorrow}
\author[1]{Brian Clark}
\author[1]{Liran Tal}
\author[2,1]{Luke Tu}
\affil[1]{Snyk}
\affil[2]{Brown University}

\date{October 12, 2026}

\begin{document}

\maketitle

\begin{abstract}
Software composition analysis (SCA) tools report far more vulnerable dependencies than teams fix: in Snyk Open Source production data, 94\% of vulnerability findings have a known fixed version, yet only about 10\% are resolved. Agentic coding tools could close this gap, but no benchmark evaluates them on vulnerability-driven upgrades in real, runnable projects across ecosystems, verified independently of the agent. We present \bench{}: 307 upgrade tasks covering 2{,}066 fixable vulnerabilities in 43 repositories across six package managers, matched to Snyk Open Source production vulnerability distributions. An upgrade passes only if, in a fresh container, dependencies resolve, a rescan shows the targeted vulnerabilities fixed, the project builds, and every previously passing test still passes. A non-agentic baseline that only applies Snyk's recommended version passes 46.9\% of upgrades; with the same model (Claude Sonnet 5.5), Snyk's \remy{} passes 90.2\%, \aided{} 77.5\% and \unaided{} 59.6\%. The agents fail in opposite ways: Claude Code mostly leaves the vulnerability in place, whereas most of \remy{}'s failures break the build or tests. An SCA scanner is a key tool for agents: Snyk's guidance raises Claude Code's PASS rate by 17.9 percentage points; without it, Claude Code often settles on a still-vulnerable version. Yet knowing the target is not enough: given the same Snyk findings, \remy{}'s purpose-built harness passes 12.7 points more than \aided{}. Major framework migrations remain hard for all harnesses. Because the agents keep projects working at similar rates, evaluating remediation quality in dependency upgrades requires verifying the fix, the build and the tests together.
\end{abstract}

\section{Introduction}
\label{sec:intro}

Software composition analysis (SCA) scanners such as Snyk Open Source~\cite{snykopensource} flag vulnerabilities in open-source dependencies, and the fix is usually an upgrade: in Snyk production data, 94.0\% of findings have a known fixed version, yet only about 10\% are resolved (Section~\ref{sec:calibration}). The bottleneck is upgrading \emph{safely}. Of 306 dependency-update pull requests that bots such as Dependabot opened in Maven projects and developers rejected after discussion, 51.3\% involved incompatibility concerns (compile, test or CI failures, dependency conflicts)~\cite{zhang2023coral}. In Snyk's data, about a quarter of fixes need a major-version jump and most vulnerable packages arrive transitively (Section~\ref{sec:calibration}), and even minor and patch releases often contain breaking changes~\cite{raemaekers2017semver,ochoa2022breakingbad}. Cautious upgrades do not solve the problem: non-major upgrades of direct dependencies fix only 25.71\% of CVEs~\cite{zhang2023coral}. An upgrade that removes the vulnerability but breaks the build or tests is not a fix.

Agentic coding tools could do this work end to end, but no benchmark measures whether they do it reliably. That requires (i)~real repositories with working builds and tests, (ii)~tasks driven by real vulnerability findings with known fix paths, and (iii)~verification, independent of the agent, that the vulnerability is gone \emph{and} the project still works. Existing benchmarks cover at most two (Section~\ref{sec:related}): breaking-update benchmarks~\cite{reyes2024bump,luo2026depbench,yang2026deprepair} are not driven by vulnerabilities, and CORAL~\cite{zhang2023coral} scores against its own CVE-to-version mapping rather than a production scan.

We introduce \bench{} to fill this gap. Each task, an \emph{upgrade}, pairs a real repository, pinned at a vulnerable commit, with one upgrade recommended by a Snyk scan. The simplest fix is a version bump, which raises the dependency's version in the manifest or lockfile and changes nothing else; harder ones also need code changes or dependency conflict resolution. A verifier re-checks the agent's work in a fresh container with four \emph{gates}: dependencies resolve, a rescan shows the targeted vulnerabilities fixed, the project builds, and every previously passing test still passes. Only an attempt that passes all four is a PASS.

We use \bench{} to evaluate four harnesses on the same 307 upgrades. The \pmbump{}, the non-agentic baseline, bumps the package to Snyk's recommended version with the project's own package manager, and changes nothing else. The three agentic harnesses use the same model, Claude Sonnet 5.5: Snyk's \remy{}, and the general-purpose Claude Code~\cite{claudecode} run with the Snyk scan and recommended upgrade (\aided{}) and without them (\unaided{}). The baseline passes 144 upgrades, a lower bound on what mechanical upgrading can solve (Section~\ref{sec:failures}). On these \emph{Snyk-advice-solvable} upgrades, an agent can only keep or lose an easy fix; on the other 163, it must do more than bump a version, such as adapting code to breaking changes or resolving dependency conflicts.

Two design choices let \bench{} tell these harnesses apart. First, because each task starts from a vulnerability, \bench{} evaluates remediation directly and can therefore measure \emph{avoidance}, an agent leaving the vulnerability in place rather than taking a risky upgrade. Breaking-update benchmarks cannot detect avoidance, yet it is Claude Code's most common failure (Section~\ref{sec:failures}). Second, a PASS requires both a remediated vulnerability and a working project. Evaluated on dependency resolution, build and tests alone, the agents perform similarly; only the remediation check distinguishes them.

This paper makes three contributions:
\begin{itemize}
  \item \textbf{\bench{}, a dataset grounded in production data:} 43 repositories, selected to match Snyk production vulnerability distributions from 99 qualifying candidates with a reproducible install, build and test baseline, yielding 307 upgrade tasks over 2{,}066 fixable vulnerability instances, from bumps that need no code changes to major migrations.
  \item \textbf{A verifier that does not trust the agent}, applying the four gates in a separate container, while the agent runs in an isolated, production-like sandbox with leakage controls.
  \item \textbf{A same-model comparison} of a purpose-built remediation engine and a general-purpose coding agent, with per-gate results, paired statistics, a failure analysis that separates avoidance from breakage, and the \pmbump{} baseline.
\end{itemize}

Our main findings are:
\begin{itemize}
  \item \textbf{Only the vulnerability check separates the agents.} On dependency resolution, build and tests alone, the three agents perform similarly (92.8--93.8\%); with the rescan, \remy{} passes 90.2\% of upgrades, \aided{} 77.5\% and \unaided{} 59.6\%, a ranking that holds per repository (Section~\ref{sec:headline}).
  \item \textbf{An SCA scanner is a key tool for agents fixing vulnerabilities.} Snyk's recommended upgrade applied by the \pmbump{} passes 46.9\% of upgrades and Claude Code without Snyk 59.6\%; together, as \aided{}, they pass 77.5\%. Without an SCA scan, Claude Code regresses on 20 of the 144 upgrades the \pmbump{} solves, 15 of them by choosing a version Snyk still flags: web search alone does not reliably find a safe version (Section~\ref{sec:naive}).
  \item \textbf{Knowing the target is not enough.} Given the same Snyk findings and model, \remy{}'s purpose-built harness passes 12.7 percentage points more than \aided{}, mostly on the 163 upgrades the \pmbump{} fails, where \remy{} passes 82.2\%, \aided{} 61.3\% and \unaided{} 36.2\% (Sections~\ref{sec:headline} and~\ref{sec:naive}).
  \item \textbf{The agents fail in opposite ways.} Claude Code mostly stops short and fails silently: the vulnerability remains in 107 of \unaided{}'s 124 failures, 70 of which it claimed it fixed, and in 48 of \aided{}'s 69. \remy{} usually lands the upgrade, so 21 of its 30 failures break the build or tests, where the project's own checks catch them (Section~\ref{sec:failures}).
  \item \textbf{Major framework migrations remain hard.} PASS rates fall from 96\%, 88\% and 70\% on non-major upgrades to 81\%, 61\% and 42\% on major ones for \remy{}, \aided{} and \unaided{}; half of \remy{}'s major PASSes edit application source, and 23 upgrades, mostly Spring, Spring Boot and Dropwizard migrations, defeat all three agents (Section~\ref{sec:hard}).
\end{itemize}

\section{Related Work}
\label{sec:related}

Existing upgrade benchmarks check that a project still works after an upgrade but not that a vulnerability was remediated, and the one vulnerability-driven evaluation covers a single ecosystem. Table~\ref{tab:related} compares the closest benchmarks on the checks they use to score an attempt.

\begin{table}[!t]
\centering
\caption{Dependency-upgrade and remediation benchmarks compared with \bench{} on the checks used to score an attempt. \checkmark{} = scored; $\circ$ = partial, e.g.\ checked only within another step or against the authors' own CVE mapping (CORAL); Vuln = rescan with a production SCA scanner; $^\dagger$ = held-out tests hidden from the agent. BUMP defines no repair scoring; LADU builds and tests in its loop but scores only overlap with manual upgrades.}
\label{tab:related}
{\small
\setlength{\tabcolsep}{4pt}
\renewcommand{\arraystretch}{1.35}
\begin{tabular}{@{}l>{\raggedright}p{2.6cm}>{\raggedright}p{2.6cm}>{\raggedright\arraybackslash}p{2.0cm}ccccc@{}}
\toprule
Benchmark & Ecosystems & Size & Task trigger & \shortstack{Real\\projects} & \shortstack{Depen-\\dencies} & Build & Tests & Vuln \\
\midrule
BUMP~\cite{reyes2024bump} & Maven & 153 repositories, 571 upgrades & Breaking update & \checkmark & $\circ$ & \checkmark & \checkmark & \\
\addlinespace[5pt]
BigBag~\cite{reyes2026bigbag} & Maven & 157 upgrades & Breaking update & \checkmark & $\circ$ & \checkmark & \checkmark & \\
\addlinespace[5pt]
DepRepair~\cite{yang2026deprepair} & Maven, npm, Cargo, PyPI & 95 upgrades & Major migration & \checkmark & & $\circ$ & $\circ$ & \\
\addlinespace[5pt]
DepBench~\cite{luo2026depbench} & npm/Yarn, Maven, Go, Cargo, Python & 203 upgrades & Bot upgrade PR & \checkmark & $\circ$ & $\circ$ & \checkmark$^\dagger$ & \\
\addlinespace[5pt]
CORAL~\cite{zhang2023coral} & Maven & 301 repositories, 750 unique CVEs & CVE & \checkmark & $\circ$ & \checkmark & \checkmark & $\circ$ \\
\addlinespace[5pt]
REQBench~\cite{lei2025pcreq} & pip & 34 repositories, 2{,}095 upgrades & Library upgrade & \checkmark & $\circ$ & $\circ$ & $\circ$ & \\
\addlinespace[5pt]
LADU~\cite{tawosi2025ladu} & Maven & 3 synthetic repositories & Major upgrade & & & & & \\
\midrule
\textbf{\bench{}} & Maven, Gradle, npm, Yarn, pip, Poetry & 43 repositories, 307 upgrades, 2{,}066 vulnerabilities & Vulnerability fix path & \checkmark & \checkmark & \checkmark & \checkmark & \checkmark \\
\bottomrule
\end{tabular}}
\end{table}

\paragraph{Breaking dependency updates.}
Benchmarks of breaking dependency updates judge an upgrade by whether the project still builds and passes its tests. BUMP~\cite{reyes2024bump} collects 571 reproducible breaking updates from 153 Maven projects, each a one-line version bump of a declared dependency that fails the build. BUMP defines no repair scoring of its own; its marks in Table~\ref{tab:related} reflect the \texttt{mvn clean test} run that reproduces each breaking update. Byam~\cite{reyes2025byam} repairs BUMP's compilation errors with LLM prompts, and BigBag~\cite{reyes2026bigbag} has an agent generate reusable AST transformation rules for them, on a subset of 157 BUMP upgrades. Tests generated by BreakGuard~\cite{raj2026breakguard} to pass before and fail after an update detect at best 30.3\% of BUMP breaking updates, almost all crashes rather than behavioural changes.

Recent benchmarks apply the same check to LLM repair across ecosystems. DepRepair~\cite{yang2026deprepair}\footnote{Yang et al.\ name their benchmark ``DepBench'' and their method DepRepair; we use ``DepRepair'' for the benchmark to avoid confusion with the benchmark of Luo et al.~\cite{luo2026depbench}.} curates 95 instances of 10 well-known major migrations across Maven, npm, Cargo and PyPI; all its methods, including the agents, generate patches from about ten pre-scoped files without an in-loop build or test environment. DepBench~\cite{luo2026depbench} mines 203 tasks from bot pull requests that needed human source or test changes, across five ecosystems, and hides its held-out upgrade tests from the agent; its best configuration solves 51.2\% of tasks, and changing only the harness moves results by up to 16.3 points. PCREQ~\cite{lei2025pcreq} infers compatible requirements for Python library upgrades to every later version. On its REQBench dataset, LLM baselines solve only 11--24\% of the 406 scenarios in which a plain \texttt{pip} upgrade crashes at runtime, yet fail 9--13\% of those \texttt{pip} already handles: the trade-off between fixing hard cases and breaking easy ones that \bench{} measures against its \pmbump{} (Section~\ref{sec:naive}). None of these benchmarks starts from a vulnerability or checks that one was removed, so none can observe an agent that leaves it in place.

\paragraph{Vulnerability remediation.}
To our knowledge, the only prior evaluation that starts from vulnerable dependencies is CORAL~\cite{zhang2023coral}, and it covers only the Java (Maven) ecosystem. It uses the Z3 SMT solver to select compatible versions for vulnerable dependencies in 301 Maven projects and reports fixing 87.56\% of CVE occurrences. It evaluates a solver rather than agents, never edits application code, and scores all tools against the authors' own NVD-derived CVE-to-version mapping, which CORAL also optimises against, rather than a production SCA scan.

\paragraph{What \bench{} adds.}
\bench{} tasks start from vulnerability remediation: several releases often fix a vulnerability, but the agent must remediate the vulnerability while keeping the build and tests intact. Taking the smallest fixing release does not guarantee this: security fixes are often patch or minor releases, which semantic versioning promises are backward compatible, but this promise often fails. On Maven Central, 35.7\% of minor and 23.8\% of patch releases contain at least one breaking change~\cite{raemaekers2017semver}; a later replication found that 83.4\% of upgrades conform to semantic versioning, yet 20.1\% of non-major upgrades still introduce breaking changes, although only 7.9\% of sampled clients would actually break~\cite{ochoa2022breakingbad}. In Go, Li et al.~\cite{li2023gosvi} found that 86.3\% of upgrades comply with semantic versioning, yet 28.6\% of non-major upgrades introduce breaking changes. Even a minimal, in-range upgrade can therefore break a consumer, and adapting code to a given library version is itself hard for models: the largest enterprise models reach only 48--51\% baseline success on GitChameleon~2.0's~\cite{misra2026gitchameleon} 328 version-conditioned Python problems, and LibEvoBench~\cite{cipollone2026libevobench}, which tests whether models use the correct API for a given version of PyTorch, NumPy or SciPy, finds them largely version-oblivious: stating the target version in the prompt does not improve accuracy.

No existing benchmark measures this joint task on real-world projects across multiple ecosystems. \bench{} checks dependency resolution, the targeted vulnerabilities, the build and the tests together in six package managers (Table~\ref{tab:related}), and includes transitive-only fixes, which BUMP's one-line bumps of declared dependencies exclude. As SWE-bench~\cite{jimenez2024swebench} did for GitHub issues, it evaluates on real repositories, with agents that build and test in the loop. LADU~\cite{tawosi2025ladu} also compiles and tests in its loop while adapting a Java project to a chosen version of an industrial framework, but is evaluated on three synthetic repositories by overlap with manual upgrades. Finally, \bench{} reports per-gate results, a first-failed-gate breakdown and repository-clustered significance tests, which none of BUMP, CORAL, REQBench, DepRepair or DepBench report.

\section{The \bench{} Dataset}
\label{sec:dataset}

\bench{} is shaped by Snyk Open Source production data, which determined the ecosystems we mined and the vulnerability profile we aimed for. From the mined open-source candidates, we kept repositories that build reproducibly, pass more than 85\% of their tests and contain at least one vulnerability, and selected those that most closely match the production distributions.

\subsection{Grounding the Dataset in Snyk Production Data}
\label{sec:calibration}

The target distributions come from Snyk Open Source findings from April to September 2026: about 300{,}000 organisations and 459 million unique \emph{issue--project pairs}. A Snyk project is one monitored manifest, and a repository can contain several. Because each project must be fixed separately, a vulnerability counts once for every project it affects. In this data, 94.0\% of findings have a known fixed version, and for 93.7\% Snyk also finds an upgrade path to it within the project, yet 89\% are still open, about 10\% resolved and 1\% ignored. Four Snyk Open Source production observations shaped \bench{}:
\begin{itemize}
  \item \textbf{Package-manager coverage.} Java (62.7\%), JavaScript (22.1\%) and Python (9.9\%) account for almost 95\% of findings; \bench{} targets the six package managers that cover about 92\% of pairs (Maven, Gradle, npm, Yarn, pip and Poetry).
  \item \textbf{Direct versus transitive.} In a 1\% sample of projects (4.2 million findings), most findings are transitive: the vulnerable package is declared in the manifest for only 14.0\% of Maven, 19.4\% of Gradle, 34.6\% of npm, 22.1\% of Yarn and 15.3\% of Poetry findings. The exception is pip (55.5\%), because \texttt{requirements.txt} files often list the full resolved set.
  \item \textbf{Severity.} High (47.5\%) and medium (40.8\%) findings dominate; critical is 7.1\% and low 4.6\%.
  \item \textbf{Fix difficulty.} The smallest semantic-versioning jump to a fixed release splits patch/\allowbreak\mbox{minor}/\allowbreak major 40/33/26\% in Java, 27/48/25\% in JavaScript and 21/50/29\% in Python: about a quarter of fixes need a major upgrade in every language.
\end{itemize}

\subsection{Candidate Repository Collection}
\label{sec:sources}

Candidates come from five sources covering all six package managers (counts after deduplication):
\begin{itemize}
  \item \textbf{BUMP projects (Maven, 133 candidates)}~\cite{reyes2024bump}, pinned at the parent of a breaking-update commit, where the project still builds and passes its tests.
  \item \textbf{Popular Maven projects (34):} top-starred Maven projects on GitHub, pinned at their last commit before 21 May 2022. This mirrors CORAL's setup of popular Maven projects at its snapshot date~\cite{zhang2023coral}, and leaves them old enough to contain open vulnerabilities.
  \item \textbf{Popular open-source projects (274):} curated applications, frameworks, starters and tutorials (Gradle, npm, Yarn, pip, Poetry). Well-maintained projects often have no open vulnerabilities at their latest commit, so many are pinned at an older one.
  \item \textbf{GitHub searches for vulnerable packages (280):} latest commits of manifests that pin known-vulnerable versions of frequently vulnerable packages (e.g.\ \texttt{axios}, \texttt{urllib3}), plus repository searches for less-represented package managers.
  \item \textbf{Gradle submodules (46):} modules of large multi-project builds (at the parent's commit), scanned separately so that one task does not span hundreds of modules.
\end{itemize}

We scan each candidate with \texttt{snyk test -{}-json} and record, per unique issue--project pair, severity, fixability, direct or transitive status and fix difficulty, the semantic-versioning jump from the installed version to the earliest fixed release. Some repositories were collected at several commits, giving 805 candidates from 721 repositories. We kept the most vulnerable commit of each, which leaves 767 candidates: one per repository plus the 46 Gradle submodules.

\subsection{Filtering for Reproducible and Vulnerable Baselines}
\label{sec:filtering}

\begin{figure}[htbp]
\centering
\includegraphics[width=\textwidth]{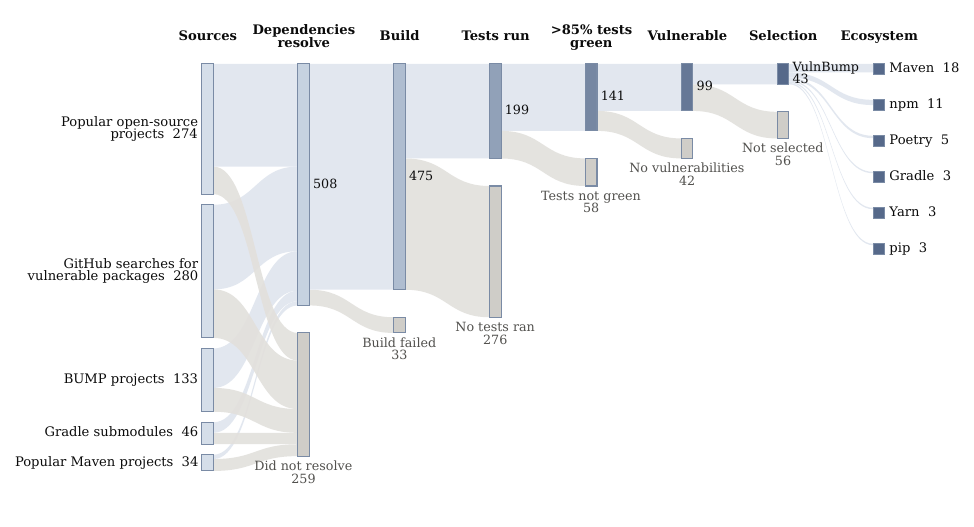}
\caption{Dataset selection: the 767 deduplicated candidates through the five filters (passing on top, dropped in grey), and the 43 selected repositories by package manager.}
\label{fig:funnel}
\end{figure}

So that any later breakage can be attributed to the upgrade, we keep only candidates with a reproducible baseline: the unmodified project must install, build and pass its tests from a clean checkout in its ecosystem's image (JDK~17, Node~20 or Python~3.11; Section~\ref{sec:sandbox}). Each step had a 30-minute limit, and each candidate had to pass five filters (Figure~\ref{fig:funnel}):
\begin{enumerate}
  \item \textbf{Dependencies resolve:} every declared dependency can be downloaded and installed with the project's own package manager (e.g.\ \texttt{mvn dependency:resolve}, \texttt{pip install}), without private registries (508 of 767 pass).
  \item \textbf{Build:} Maven and Gradle projects compile; other ecosystems have no build step (475).
  \item \textbf{Tests run:} at least one test is discovered and executed (199).
  \item \textbf{>85\% tests green} on the unmodified project (141).
  \item \textbf{Vulnerable:} Snyk Open Source reports at least one vulnerability (99).
\end{enumerate}

The largest drop is at the third filter: many projects have no tests that run in a clean environment, and most pip projects need external services such as Postgres or Redis. Earlier failures were mostly ecosystem-specific: multi-module resolution, private registries and plugins (Maven, Gradle), peer-dependency conflicts (npm), and lockfile or Python-version mismatches (Poetry). This leaves 99 qualifying candidates from 97 repositories (55 Maven, 9 Gradle, 19 npm, 8 Yarn, 3 pip and 5 Poetry).

\subsection{Repository Selection}
\label{sec:selection}

From the 99 qualifying candidates we selected 43 repositories that match Snyk production data in direct versus transitive share, vulnerabilities per repository, severity and fix difficulty, while giving each package manager enough cases to report on rather than weighting strictly by production share. The final \bench{} dataset comprises 18 Maven, 3 Gradle, 11 npm, 3 Yarn, 3 pip and 5 Poetry repositories (21 Java, 14 JavaScript, 8 Python), with a median of 32 unique vulnerabilities (range 1--252) and 88 passing tests; 19 repositories have at least 1{,}000 GitHub stars.

\begin{figure}[tb]
\centering
\includegraphics[width=\textwidth]{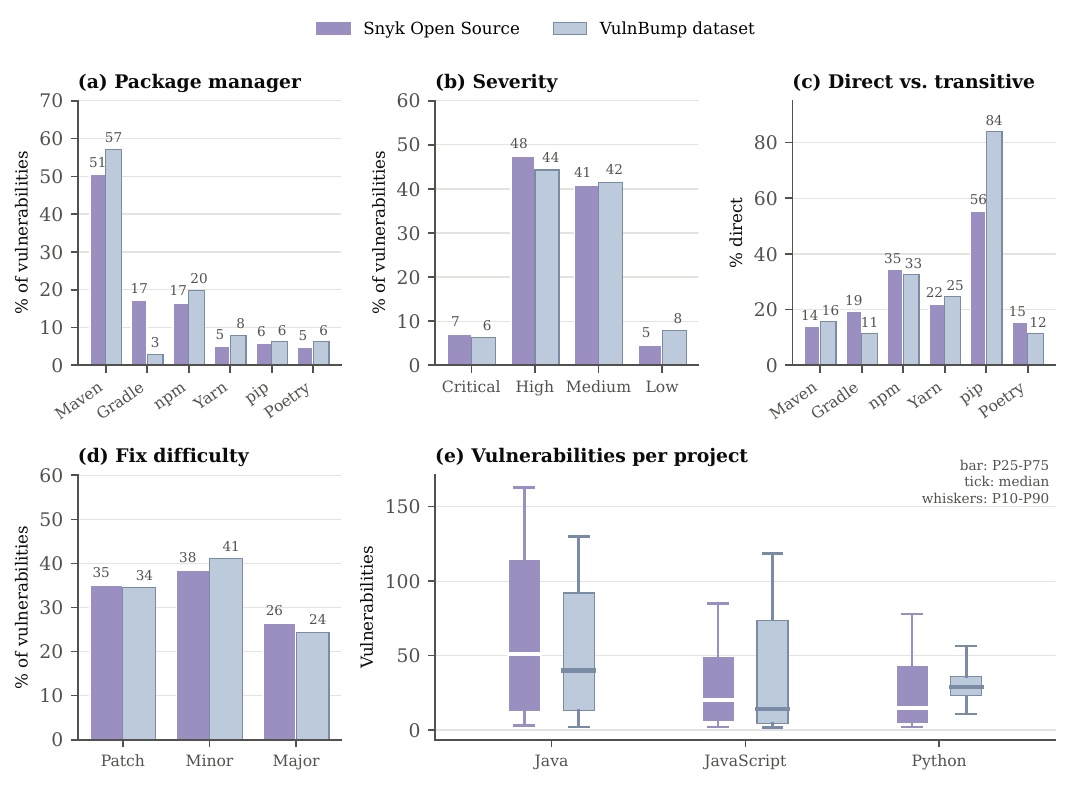}
\caption{Snyk Open Source production data (April--September 2026) versus \bench{} dataset distributions (43 repositories). Production fix difficulty covers Java, JavaScript and Python findings. Production direct shares come from a 1\% random sample of projects.}
\label{fig:distributions}
\end{figure}

The dataset matches production within a few points on severity, fix difficulty, and on package-manager share except for Gradle, and on direct share for Maven, npm, Yarn and Poetry (Figure~\ref{fig:distributions}). Vulnerabilities per project span a similar range for Java and JavaScript, while the Python repositories cluster around a higher median.

Two deviations from Snyk Open Source production distributions remain. First, Gradle is under-represented (3\% of vulnerabilities versus 17\%), with a lower direct share (11\% versus 19\%): most Gradle candidates failed the filters, nearly half of them at dependency resolution, and the largest qualifying ones were too large to evaluate (Section~\ref{sec:limitations}). Second, pip vulnerabilities are mostly direct (84\% versus 56\%), because only three pip repositories passed the baseline filtering.

\section{Remediation Task and Fix Verification}
\label{sec:design}

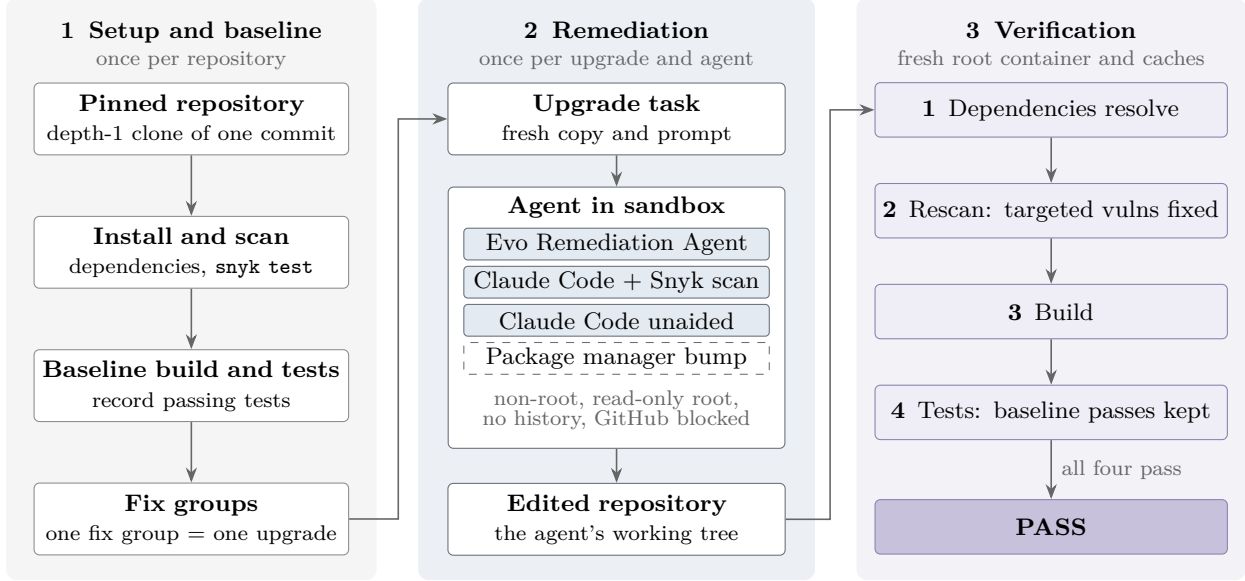
\begin{figure}[htbp]
\centering
\begin{tikzpicture}[
  font=\footnotesize,
  >={Stealth[length=2.0mm]},
  flow/.style={->, draw=black!60, line width=0.6pt},
  base/.style={rounded corners=2pt, align=center, inner sep=3pt, fill=white},
  sbox/.style={base, draw=black!45, minimum height=0.95cm, text width=3.95cm},
  rbox/.style={base, draw=greyblue!55!black, minimum height=0.95cm, text width=4.25cm},
  gate/.style={base, draw=accent!90!black, fill=accent!16, minimum height=0.72cm, text width=4.45cm},
  pass/.style={base, draw=accent!90!black, fill=accent!55, minimum height=0.72cm, text width=4.45cm, font=\footnotesize\bfseries},
  chip/.style={rounded corners=1.5pt, draw=greyblue!55!black, fill=greyblue!40, inner sep=2pt, text width=3.9cm, align=center, minimum height=0.42cm},
  head/.style={anchor=north, align=center, inner sep=0pt},
  sub/.style={font=\scriptsize, text=black!60},
]
\def\T{0.0}\def\B{-7.70}
\begin{scope}[on background layer]
\fill[rounded corners=4pt, black!4]     (0.00,\T) rectangle (4.90,\B);
\fill[rounded corners=4pt, greyblue!28] (5.45,\T) rectangle (10.70,\B);
\fill[rounded corners=4pt, accent!12]   (11.25,\T) rectangle (16.40,\B);
\end{scope}
\node[head] at (2.45,-0.30)  {\textbf{1\enspace Setup and baseline}\\[-1pt]{\scriptsize\color{black!60} once per repository}};
\node[head] at (8.075,-0.30)  {\textbf{2\enspace Remediation}\\[-1pt]{\scriptsize\color{black!60} once per upgrade and agent}};
\node[head] at (13.825,-0.30) {\textbf{3\enspace Verification}\\[-1pt]{\scriptsize\color{black!60} fresh root container and caches}};

\node[sbox] (repo) at (2.45,-1.60) {\textbf{Pinned repository}\\{\scriptsize depth-1 clone of one commit}};
\node[sbox, below=0.8cm of repo] (scan) {\textbf{Install and scan}\\{\scriptsize dependencies, \texttt{snyk test}}};
\node[sbox, below=0.8cm of scan] (test) {\textbf{Baseline build and tests}\\{\scriptsize record passing tests}};
\node[sbox, below=0.8cm of test] (fix)  {\textbf{Fix groups}\\{\scriptsize one fix group = one upgrade}};
\draw[flow] (repo) -- (scan);
\draw[flow] (scan) -- (test);
\draw[flow] (test) -- (fix);

\node[rbox] (task) at (8.075,-1.60) {\textbf{Upgrade task}\\{\scriptsize fresh copy and prompt}};
\node[inner sep=0pt] (atitle) at (8.075,-2.77) {\textbf{Agent in sandbox}};
\node[chip, below=0.12cm of atitle] (c1) {\remy{}};
\node[chip, below=0.07cm of c1] (c2) {Claude Code + Snyk scan};
\node[chip, below=0.07cm of c2] (c3) {Claude Code unaided};
\node[chip, below=0.07cm of c3, fill=white, dashed] (c4) {Package manager bump};
\node[sub, below=0.07cm of c4, align=center] (anote) {non-root, read-only root,\\[-2pt]no history, GitHub blocked};
\begin{scope}[on background layer]
  \node[base, draw=greyblue!55!black, fit=(atitle)(c1)(c4)(anote), inner xsep=0.2cm, inner ysep=0.12cm] (agent) {};
\end{scope}
\node[rbox, below=0.45cm of agent] (tree) {\textbf{Edited repository}\\{\scriptsize the agent's working tree}};
\draw[flow] (task) -- (agent);
\draw[flow] (agent) -- (tree);

\node[gate] (g1) at (13.825,-1.48) {\textbf{1}\enspace Dependencies resolve};
\node[gate, below=0.6cm of g1] (g2) {\textbf{2}\enspace Rescan: targeted vulns fixed};
\node[gate, below=0.6cm of g2] (g3) {\textbf{3}\enspace Build};
\node[gate, below=0.6cm of g3] (g4) {\textbf{4}\enspace Tests: baseline passes kept};
\node[pass, below=0.78cm of g4] (pass) {PASS};
\draw[flow] (g1) -- (g2);
\draw[flow] (g2) -- (g3);
\draw[flow] (g3) -- (g4);
\draw[flow] (g4) -- node[right, sub] {all four pass} (pass);

\draw[flow] (fix.east) -| ($(fix.east)!0.5!(task.west)$) |- (task.west);
\draw[flow] (tree.east) -| ($(tree.east)!0.5!(g1.west)$) |- (g1.west);
\end{tikzpicture}
\caption{The \bench{} pipeline: (1)~Setup and baseline: the pinned repository is installed, scanned, built and tested, and every Snyk fix group becomes one upgrade; (2)~Remediation: once per upgrade, the agent edits a fresh repository copy in a sandbox, while the \pmbump{} (dashed) only applies Snyk's recommended version and refreshes the lockfile; (3)~Verification: a fresh container applies four gates to the agent's working tree; PASS requires all four.}
\label{fig:pipeline}
\end{figure}

This section defines the upgrade task, the four harnesses, the controlled sandbox in which they run, the verification gates and the metrics; Figure~\ref{fig:pipeline} shows how they fit together.

\subsection{Remediation Task}
\label{sec:task}

Each task asks an agent to apply one package upgrade (or pin) recommended by a baseline Snyk scan (\texttt{snyk test -{}-all-projects -{}-ignore-policy}) of the installed, pinned repository, which must first install, scan, build and pass at least one test at evaluation time. Snyk groups fixable vulnerabilities into \emph{fix groups}: one upgrade or pin in one manifest, with the vulnerabilities it fixes. Vulnerabilities without a fix path are never counted against an agent.

We call each fix group an \emph{upgrade}; it is the unit of a \bench{} task, so a repository whose 50 fixable vulnerabilities require upgrading 20 packages yields 20 upgrades. Within an upgrade, we count \emph{fixable vulnerability instances}: issue--project pairs in the baseline scan for which Snyk offers an upgrade or pin. If a vulnerability is reachable through several paths, each fixed by a different upgrade, it counts as fixed only when every one of those upgrades removes its path (Listing~\ref{lst:snyk}).

\begin{lstlisting}[caption={Baseline \texttt{snyk test} output for one \bench{} repository (expressjs/generator). Its seven vulnerabilities form four upgrades, one per ``Upgrade'' block; the \texttt{mkdirp} upgrade fixes a transitive vulnerability (\texttt{mkdirp > minimist}), and \texttt{SNYK-JS-\allowbreak MINIMIST-\allowbreak 2429795} is reachable through both the \texttt{minimist} and \texttt{mkdirp} upgrades, so it counts as fixed only if both packages are bumped.},label={lst:snyk},literate={✗}{{$\times$}}1,basicstyle=\ttfamily\fontsize{7.5}{8.8}\selectfont,breakatwhitespace=true]
Tested 9 dependencies for known issues, found 7 issues, 8 vulnerable paths.

Issues to fix by upgrading:

  Upgrade ejs@2.6.2 to ejs@3.1.10 to fix
  ✗ Improper Control of Dynamically-Managed Code Resources [Medium Severity][SNYK-JS-EJS-6689533] in ejs@2.6.2
    introduced by ejs@2.6.2
  ✗ Remote Code Execution (RCE) [High Severity][SNYK-JS-EJS-2803307] in ejs@2.6.2
    introduced by ejs@2.6.2

  Upgrade minimatch@3.0.4 to minimatch@3.1.3 to fix
  ✗ Regular Expression Denial of Service (ReDoS) [Medium Severity][SNYK-JS-MINIMATCH-3050818] in minimatch@3.0.4
    introduced by minimatch@3.0.4
  ✗ Regular Expression Denial of Service (ReDoS) [High Severity][SNYK-JS-MINIMATCH-15309438] in minimatch@3.0.4
    introduced by minimatch@3.0.4
  ✗ Inefficient Algorithmic Complexity [High Severity][SNYK-JS-MINIMATCH-15353389] in minimatch@3.0.4
    introduced by minimatch@3.0.4

  Upgrade minimist@1.2.5 to minimist@1.2.6 to fix
  ✗ Prototype Pollution [Low Severity][SNYK-JS-MINIMIST-2429795] in minimist@1.2.5
    introduced by minimist@1.2.5 and 1 other path(s)

  Upgrade mkdirp@0.5.1 to mkdirp@0.5.5 to fix
  ✗ Prototype Pollution [Low Severity][SNYK-JS-MINIMIST-2429795] in minimist@1.2.5
    introduced by minimist@1.2.5 and 1 other path(s)
  ✗ Prototype Pollution [Medium Severity][SNYK-JS-MINIMIST-559764] in minimist@0.0.8
    introduced by mkdirp@0.5.1 > minimist@0.0.8

Package manager:   npm
Target file:       package-lock.json
\end{lstlisting}

Each upgrade is attempted from a fresh copy of the pinned repository and scored on its own (\emph{single-upgrade} mode), so one upgrade's failure cannot affect another.

\subsection{Harnesses}
\label{sec:agents}

\paragraph{Package manager bump.} The \pmbump{}, our non-agentic baseline, applies Snyk's recommended upgrade and nothing else: it uses no model, edits no code or tests, and never tries another version. It sets the named package's version with the ecosystem's own mechanism (\texttt{npm install pkg@ver}, \texttt{yarn add}, \texttt{poetry add}, or the version in \texttt{requirements.txt}, \texttt{pom.xml} or the Gradle build), and pins an undeclared package the same way, e.g.\ with an explicit Maven \texttt{<version>} or a Gradle \texttt{constraints} entry. A pin is \emph{managed} when it overrides a version set by a Maven parent or BOM or by Spring's dependency-management plugin. If the lockfile refresh fails, so does the dependency gate; the only fallback is retrying npm with \texttt{-{}-legacy-peer-deps}. Python packages are pinned exactly (\texttt{==}); pinning with \texttt{>=} changes the outcome of 2 of the 60 Python upgrades and leaves the total unchanged.

\paragraph{Agentic harnesses.} All three agents use the same model (\texttt{claude-sonnet-5-5} on Vertex AI) and a limit of 110 agent turns per upgrade. They differ in harness, prompt, tools and Snyk information, so we compare complete systems as a user would get them, rather than ablating one component.
\begin{description}
  \item[\remy{}] is Snyk's agentic remediation engine, available through the Snyk CLI and a cloud service. It leverages \emph{Snyk intelligence}, which evolved from Snyk's deterministic SCA fix engine for pull requests~\cite{snykfixdocs}. We invoke the CLI~\cite{snykcli} once per upgrade with the IDs of the vulnerabilities that upgrade fixes:
\begin{lstlisting}
snyk fix --agentic --sca --auto-approve --prepare-env --issue-ids <ids>
\end{lstlisting}
With \texttt{-{}-auto-approve}, it applies every fix without prompting, including upgrades Snyk rates as likely to break the build, which a user could otherwise skip. It works from the same Snyk findings as \aided{}, runs its own scan, the build and tests, and repairs errors before finishing.
  \item[\aided{}] is Claude Code 2.1.289, prompted with the project's test command and the upgrade's ``\texttt{Upgrade X to Y to fix}'' block from the baseline report, which names the target version. It may run Snyk Open Source~\cite{snykopensource} (\texttt{snyk test}).
  \item[\unaided{}] gets the same prompt without the target version (``\texttt{Upgrade X to fix these vulnerabilities:}'') and no Snyk access, so it must find a fixed version itself, e.g.\ from OSV~\cite{osv}.
\end{description}

\subsection{Sandbox and Leakage Controls}
\label{sec:sandbox}

The agent runs in a restricted container modelled on \remy{}'s production sandbox: a non-root user, a read-only filesystem except for the repository and a scratch directory, and 24\,GB of memory. The ecosystem image (JDK~17 with Maven or Gradle, Node~20, or Python~3.11 with pip or Poetry) adds Node.js, a shell, \texttt{jq} and \texttt{yq}, and the project's dependencies are installed beforehand. Package registries are reachable (Maven Central through a read-through cache); Gradle wrapper distributions are pre-seeded because GitHub is blocked. Claude Code runs headless, without permission prompts, with its standard tools~\cite{claudecodetools} (shell; file read, edit and write; web fetch and web search). \remy{} has its own tools to read, edit and run commands, and no web tool.

Three controls limit leakage of the eventual fix:
\begin{itemize}
  \item \textbf{No commit history.} Each repository is a depth-1 clone of the pinned commit, with no remote, refs or tags.
  \item \textbf{No GitHub.} GitHub and mirrors that serve its repository files (jsDelivr, raw.githack, Sourcegraph) are blocked, and Claude Code has matching \texttt{WebFetch} deny rules and no \texttt{gh} CLI. Two repositories whose setup fetches from GitHub are exempt from the host block; only the \texttt{WebFetch} rules apply there.
  \item \textbf{No Snyk for \unaided{}.} It gets no Snyk CLI or token: \texttt{snyk.io}, \texttt{security.\allowbreak snyk.io}, \texttt{api.snyk.io} and \texttt{app.snyk.io} are blocked, and links to \texttt{security.\allowbreak snyk.io} are removed from its prompt.
\end{itemize}
The rest of the web, including changelogs and migration guides hosted outside GitHub, stays open; release notes hosted on GitHub are blocked with the rest of GitHub. Section~\ref{sec:limitations} discusses residual risks.

\subsection{Verification}
\label{sec:verification}

Every fix is verified independently, without relying on the agent's report. The verifier checks the agent's working tree as root in a fresh container, with fresh caches and output directory, so nothing the agent left outside the repository (e.g.\ PATH shims or a local Maven repository) can affect the evaluation metrics. Verification runs all four gates in order, even when an earlier gate fails:
\begin{enumerate}
  \item \textbf{Dependency resolution}: the edited manifests and lockfiles install with the strict command CI would run, so session-only workarounds (e.g.\ an unsaved install flag) do not count.
  \begin{itemize}
    \item \emph{Commands:} \texttt{mvn} \texttt{dependency:resolve}, \texttt{npm ci}, \texttt{yarn install} \texttt{-{}-frozen-lockfile}, \texttt{poetry install}, or \texttt{pip install} \texttt{-r requirements.txt} plus the project; \texttt{npm install} for the three npm repositories without a committed lockfile.
    \item \emph{Java:} Gradle resolves dependencies during the build, so its resolution failures surface at the build gate; for Maven, which settles version conflicts itself, the gate is effectively a build precondition.
    \item \emph{Fallback:} if the strict install fails, npm and Yarn repositories install leniently (\texttt{npm install -{}-legacy-peer-deps}, \texttt{yarn install}) so that the remaining gates can still run.
  \end{itemize}
  \item \textbf{Rescan}: in a post-fix \texttt{snyk test} of every target manifest, the targeted vulnerabilities count as fixed when they are no longer reachable through the upgraded package or through any newly introduced dependency path. A rescan that fails or omits a target manifest fails the gate. Both scans use \texttt{-{}-ignore-policy}, so a \texttt{.snyk} file cannot hide a vulnerability.
  \item \textbf{Build}: Maven and Gradle projects compile (\texttt{mvn compile}, \texttt{gradlew compileJava}); JavaScript projects run their TypeScript type-check (\texttt{tsc -{}-noEmit}) or build script; Python projects import their main package, catching removed modules or symbols. Seven JavaScript repositories have neither a type-check nor a build script, so only their tests catch breakage.
  \item \textbf{Tests}: every baseline-passing test still passes; a skipped or missing test is a regression. Tests are matched by name from JUnit XML (Maven, Gradle, pytest) or Jest JSON reports; for the six Mocha-based npm repositories without JUnit output, the passing count in the log must not drop below the baseline's, and a log with no count fails. Agents may adapt tests to breaking changes (renames, API migrations, imports) but must not edit tests to make them pass.
\end{enumerate}
An attempt is a \textbf{PASS} only if all four gates pass. For the \texttt{minimist} upgrade in Listing~\ref{lst:snyk}, \texttt{npm ci} must accept the edited \texttt{package-lock.json}, the rescan must no longer find \texttt{SNYK-JS-\allowbreak MINIMIST-\allowbreak 2429795} on a path through \texttt{minimist}, the build gate does not apply, and Mocha must still report at least the baseline's 139 passing tests.

\subsection{Metrics}
\label{sec:metrics}

\begin{description}
  \item[Gate pass rates.] The share of the 307 upgrades that pass each gate individually, and all four gates (the PASS rate). A gate that does not apply, such as dependency resolution for Gradle, counts as passed.
  \item[Vulnerability-fix rate.] The share of the 2{,}066 fixable vulnerability instances fixed (Section~\ref{sec:task}); an instance whose rescan failed counts as not fixed.
  \item[Per-repository (macro) rates.] Repositories contribute 1 to 46 upgrades, so pooled rates, computed over all 307 upgrades together, lean towards a few large repositories. Macro rates average per-repository rates; we report them for the PASS rate.
  \item[Paired comparisons.] For two agents on the same upgrades, we test the macro PASS-rate difference for significance with a cluster bootstrap over repositories~\cite{efron1993bootstrap,davison1997bootstrap} (10{,}000 resamples), which accounts for correlation within a repository.
  \item[Snyk-advice-solvable split.] The upgrades the \pmbump{} passes, a lower bound on what mechanical upgrading can solve (Section~\ref{sec:failures}). An agent's \emph{regressions} are upgrades the \pmbump{} passes and the agent fails; its \emph{gains}, upgrades it passes and the \pmbump{} fails. We report PASS rates on both sides.
\end{description}

\section{Results}
\label{sec:results}

All four harnesses ran on the same 307 upgrades from 43 repositories. The three agents share identical baselines: 2{,}066 fixable vulnerability instances in their baseline Snyk scans. The \pmbump{} ran 3 days later on the same pinned repositories, and its 307 upgrades pair one-to-one with the agents': 305 have the same recommended upgrade, and two the same package with a newer target version (Section~\ref{sec:data}). There were no infrastructure errors or wall-clock timeouts; four failed verification rescans (one for the \pmbump{}, one for \aided{}, two for \unaided{}) count as failures.

\subsection{Headline Results}
\label{sec:headline}

\begin{table}[htbp]
\centering
\caption{Key results on \bench{}, with Claude Sonnet 5.5 for all agents; the \pmbump{} is the non-agentic baseline. Gate and PASS rates are out of 307 upgrades. \emph{Targeted vulnerabilities fixed} is the rescan gate. Vulnerability instances are out of 2{,}066 (Section~\ref{sec:metrics}).}
\label{tab:key}
\resizebox{\textwidth}{!}{%
\begin{tabular}{lcccc}
\toprule
 & \shortstack{Package manager\\bump} & \shortstack{Claude Code\\unaided} & \shortstack{Claude Code\\+ Snyk scan} & \shortstack{Evo Remediation\\Agent} \\
\midrule
Dependencies resolve $\uparrow$ & 82.4\% & 95.4\% & 94.8\% & \textbf{100\%} \\
Targeted vulnerabilities fixed $\uparrow$ & 62.2\% & 63.5\% & 83.7\% & \textbf{97.1\%} \\
Build $\uparrow$ & 87.0\% & \textbf{98.4\%} & 98.0\% & 96.7\% \\
Tests pass $\uparrow$ & 80.5\% & 98.7\% & \textbf{99.0\%} & 95.4\% \\
\midrule
All gates pass (PASS) $\uparrow$ & 46.9\% & 59.6\% & 77.5\% & \textbf{90.2\%} \\
 & (144/307) & (183/307) & (238/307) & (277/307) \\
\midrule
Vulnerability instances fixed $\uparrow$ & 56.5\% & 72.7\% & 83.1\% & \textbf{95.9\%} \\
 & (1{,}168/2{,}066) & (1{,}503/2{,}066) & (1{,}716/2{,}066) & (1{,}982/2{,}066) \\
\bottomrule
\end{tabular}}
\end{table}

\remy{} passes every gate on 90.2\% of upgrades, 12.7 percentage points more than \aided{} and 30.6 more than \unaided{}, and fixes 95.9\% of vulnerability instances (Table~\ref{tab:key}). The Snyk scan and recommended upgrade raise Claude Code's PASS rate by 17.9 points. The \pmbump{}, which applies Snyk's recommended upgrade and nothing else, passes 46.9\% and trails every agent on every gate.

The individual gates show a trade-off. \remy{} never leaves dependencies unresolved and fixes every targeted vulnerability in 97.1\% of upgrades, but its build and test rates trail \aided{}'s by 1.3 and 3.6 percentage points. On dependency resolution, build and tests combined, the agents perform similarly (92.8\% for \remy{} and \aided{}, 93.8\% for \unaided{}). Claude Code's high build and test rates partly reflect avoidance: it often leaves the vulnerability in place, and an untouched dependency rarely breaks anything (Section~\ref{sec:failures}). The \pmbump{} falls short on both halves: it fixes the targeted vulnerabilities about as often as \unaided{} (62.2\%), but keeps the project working in only 64.2\% of upgrades, because it never repairs what the upgrade breaks.

The ranking holds across repositories and package managers. Macro PASS rates, which weight every repository equally, are 51.3\%, 68.0\%, 80.1\% and 93.1\% for the \pmbump{}, \unaided{}, \aided{} and \remy{}, and every pairwise difference between the agents is statistically significant (Section~\ref{sec:metrics}). \remy{}'s macro PASS rate is at or above both Claude Code agents in every package manager, although several have only three repositories.

\subsection{Snyk's Guidance Secures the Easy Half; Agents Separate on the Rest}
\label{sec:naive}

\begin{table}[htbp]
\centering
\caption{Agentic harnesses on 144 Snyk-advice-solvable upgrades, which the \pmbump{} passes, and on the other 163, which it fails. The split is a lower bound on what mechanical upgrading can solve (Section~\ref{sec:failures}). \emph{Regressions}: the \pmbump{} passes, the agent fails. \emph{Gains}: the agent passes, the \pmbump{} fails.}
\label{tab:naive}
{\small
\begin{tabular}{lccc}
\toprule
 & \shortstack{Claude Code\\unaided} & \shortstack{Claude Code\\+ Snyk scan} & \shortstack{Evo Remediation\\Agent} \\
\midrule
PASS on Snyk-advice-solvable (144) $\uparrow$ & 86.1\% & 95.8\% & \textbf{99.3\%} \\
\quad Regressions $\downarrow$ & 20 & 6 & \textbf{1} \\
\midrule
PASS on other upgrades (163) $\uparrow$ & 36.2\% & 61.3\% & \textbf{82.2\%} \\
\quad Gains $\uparrow$ & 59 & 100 & \textbf{134} \\
\bottomrule
\end{tabular}}
\end{table}

Neither Snyk's guidance nor an agent is enough on its own. Applied by the \pmbump{}, without an agent, Snyk's recommended upgrade passes 46.9\% of upgrades; Claude Code without Snyk passes 59.6\%; given the Snyk scan and recommended upgrade, Claude Code passes 77.5\% (Table~\ref{tab:key}). The split shows what each contributes (Table~\ref{tab:naive}); per-repository (macro) rates give the same ranking on both splits.

On the 144 Snyk-advice-solvable upgrades, what matters is knowing which version is safe. With Snyk's guidance, the agents rarely regress (\aided{} on 6 upgrades, \remy{} on 1); without it, \unaided{} regresses on 20, in 15 of them by upgrading to a version Snyk still flags. Finding a fixed version by web search, e.g.\ on OSV, proved unreliable; an SCA scan names the fixed version and can confirm the upgrade. The pattern holds across all upgrades: \unaided{} chose a still-vulnerable version 64 times, \aided{} 13 times (Table~\ref{tab:causes}). Vulnerability databases can disagree on which version fixes a vulnerability, and the rescan uses Snyk Open Source, so part of the gap may reflect such disagreement (Section~\ref{sec:limitations}).

On the other 163 upgrades, which need more than a version bump, the agent decides the outcome: PASS rates are 82.2\%, 61.3\% and 36.2\% for \remy{}, \aided{} and \unaided{}, and these upgrades supply 34 of \remy{}'s 39 extra PASSes over \aided{}.

\subsection{The Agents Fail in Opposite Ways}
\label{sec:failures}

\begin{figure}[H]
\centering
\includegraphics[width=\textwidth]{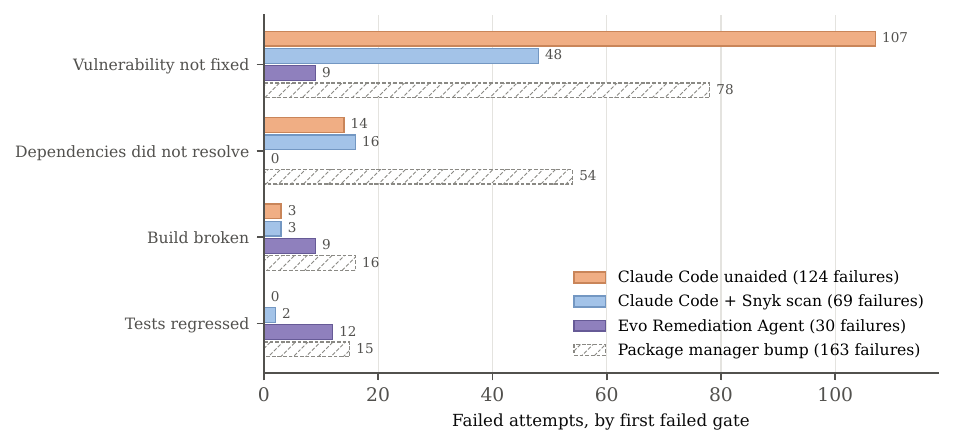}
\caption{First failed verification gate for every failed attempt (\Pmbump{}: 163 failures, \unaided{}: 124, \aided{}: 69, \remy{}: 30).}
\label{fig:failures}
\end{figure}

Claude Code mostly stops short of the fix; \remy{} lands the upgrade and sometimes breaks the code (Figure~\ref{fig:failures}). The rescan is the first failed gate in 107 of \unaided{}'s 124 failures and 48 of \aided{}'s 69, but in only 9 of \remy{}'s 30; 21 of \remy{}'s failures break the build or tests, and on 11 of these upgrades \aided{} left the vulnerability in place. Table~\ref{tab:causes} gives the root causes.

\paragraph{Claude Code fails silently.} It often avoids the risky upgrade: instead of the major version that fixes every targeted vulnerability, it settles for a patch or minor upgrade that fixes only some vulnerabilities. \unaided{} stayed within the current major version in 30 attempts, and even when told the upgrade target version, \aided{} did so in 10 (e.g.\ \texttt{spring-\allowbreak boot-\allowbreak starter-\allowbreak test} 2.7.18 instead of 3.5.15). \aided{} also pinned or excluded transitive packages instead (15) or made no change (12). In 70 of its 124 failures, \unaided{} claimed to have fixed vulnerabilities that the rescan still found. Other fixes are incomplete outside the agent's session:
\begin{itemize}
  \item \textbf{Unsaved dependency changes.} Some fixes worked in Claude Code's session but not in a clean install: in \texttt{bulletproof-\allowbreak nodejs}, \aided{} left \texttt{package.json} and \texttt{package-\allowbreak lock.json} out of sync, and in \texttt{express-\allowbreak typescript-\allowbreak starter}, Claude Code installed with \texttt{-{}-legacy-\allowbreak peer-\allowbreak deps} without saving the flag. Of the 30 Claude Code attempts that failed the strict dependency gate, 19 passed every other gate; \remy{} never failed it.
  \item \textbf{Interpreter-specific fixes.} In 15 \unaided{} and 8 \aided{} failures, the fixed release requires a newer Python than the project targets (e.g.\ \texttt{filelock} 3.20.3 requires Python~3.10); the agent locked the fix only for newer interpreters, leaving the project's own on the vulnerable version.
\end{itemize}

\begin{table}[H]
\centering
\caption{Root cause of every failed attempt, grouped by cause rather than by first failed gate, so counts can differ from Figure~\ref{fig:failures}. \emph{Vulnerability remediation}: the targeted vulnerability remains. \emph{Version compatibility}: the new versions violate a version the project pins or requires. \emph{Code compatibility}: changed API or behaviour breaks the project. \emph{Other}: a package removed from PyPI, a source build failure and a failed rescan.}
\label{tab:causes}
{\small
\setlength{\tabcolsep}{3pt}
\begin{tabular}{lcccc}
\toprule
Cause & \shortstack{Package manager\\bump} & \shortstack{Claude Code\\unaided} & \shortstack{Claude Code\\+ Snyk scan} & \shortstack{Evo Remediation\\Agent} \\
\midrule
\textbf{Vulnerability remediation} & 77 & 92 & 40 & 8 \\
\quad Upgraded to still-vulnerable version & 2 & 64 & 13 & 1 \\
\quad Still reachable via another path & 75 & 13 & 15 & 3 \\
\quad No upgrade made, or reverted & 0 & 15 & 12 & 4 \\
\midrule
\textbf{Version compatibility} & 54 & 30 & 25 & 7 \\
\quad Resolver rejects dependency versions & 25 & 14 & 16 & 0 \\
\quad Fix requires newer Python & 26 & 15 & 8 & 1 \\
\quad Dependencies need a newer compiler & 3 & 1 & 1 & 6 \\
\midrule
\textbf{Code compatibility} & 28 & 2 & 4 & 15 \\
\quad Changed API breaks build & 13 & 2 & 2 & 3 \\
\quad Class or method missing at runtime & 12 & 0 & 1 & 6 \\
\quad Changed behaviour fails tests & 3 & 0 & 1 & 6 \\
\midrule
\textbf{Other} & 4 & 0 & 0 & 0 \\
\midrule
Failed attempts & 163 & 124 & 69 & 30 \\
\bottomrule
\end{tabular}}
\end{table}

\paragraph{\remy{} fails visibly.} It attempts the upgrades Claude Code often avoids, so half its failures (15 of 30) are code incompatibilities, 13 of them on major upgrades, such as Spring~6's move from \texttt{javax.*} to \texttt{jakarta.*}. It also sometimes upgrades more than needed: in two TypeScript repositories, a broad lockfile regeneration moved transitive type definitions beyond what the project's older compiler can parse (6 failures), and its only regression came from a caret range (\texttt{agenda@\textasciicircum4.0.0}) that floated to a new major version of the MongoDB driver.

\paragraph{The \pmbump{} fails at every gate.} It always targets the recommended version but never repairs what the upgrade breaks: 54 of its 163 failures first fail at dependency resolution and 31 at the build or tests. In 75, the vulnerability stays reachable through another path, mostly in managed builds, where pinning only the named package leaves the rest of its BOM-managed family at old versions. The Snyk-advice-solvable split is therefore a lower bound on what mechanical upgrading can solve.

\subsection{Major Framework Migrations Remain Hard}
\label{sec:hard}
\label{sec:jump}

\begin{figure}[htbp]
\centering
\includegraphics[width=0.9\textwidth]{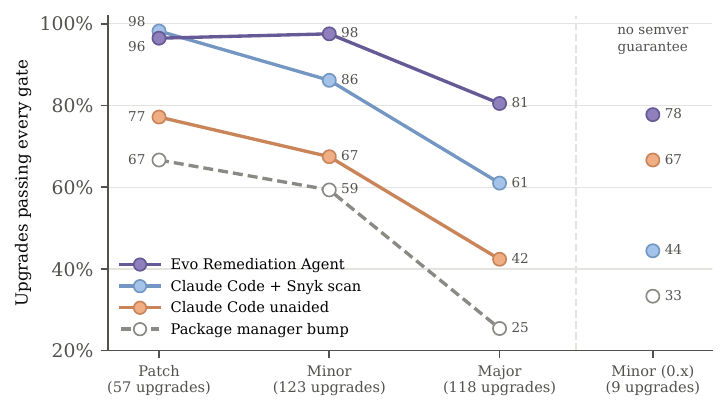}
\caption{Share of upgrades passing every gate, by the size of the recommended version jump (not the smallest fixing jump of Section~\ref{sec:calibration}). Minor jumps within 0.x versions are shown separately because semantic versioning gives no compatibility guarantee below 1.0.}
\label{fig:jump}
\end{figure}

Major upgrades separate the agents most (Figure~\ref{fig:jump}): PASS rates fall from 96\%, 88\% and 70\% on the 189 non-major upgrades to 81\%, 61\% and 42\% on the 118 major ones for \remy{}, \aided{} and \unaided{}. The \pmbump{} passes 67\%, 58\% and 25\% of patch, minor and major upgrades.

Major upgrades are where code changes are needed. \remy{} edited non-test application source in 52\% of its 95 major PASSes, against 5\% of its 182 non-major ones; Claude Code did so less often (\aided{} 32\% and 3\%, \unaided{} 24\% and 2\%). They are also where headroom remains: 23 upgrades in seven Maven and npm repositories defeated all three agents, mostly major framework migrations (Spring~5$\to$6/7, Spring Boot~2$\to$3/4, Dropwizard~2$\to$5).

Smaller jumps are not always safe: 10 of the 189 patch and minor upgrades broke the build or tests for at least one agent, e.g.\ \texttt{logback-classic} 1.2.11$\to$1.6.3 (a binary incompatibility) and \texttt{flask} 3.0.3$\to$3.1.3. This is consistent with semantic-versioning studies (Section~\ref{sec:related}) and is why \bench{} verifies every fix against the build and tests, whatever the jump.

\subsection{Transitive Vulnerabilities Are Harder for Every Harness}
\label{sec:transitive}

\begin{figure}[htbp]
\centering
\includegraphics[width=\textwidth]{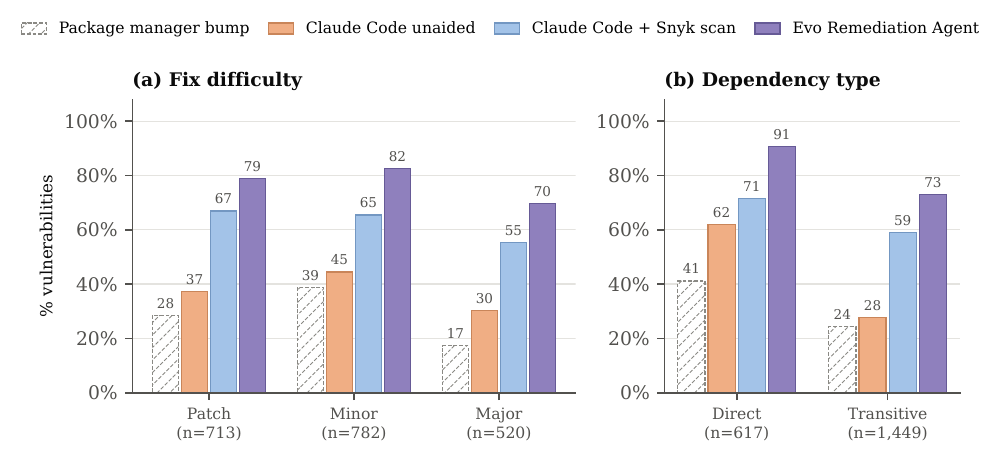}
\caption{Share of fixable vulnerabilities fixed by passing upgrades, by (a) fix difficulty (smallest semantic-versioning jump to a fixed release) and (b) direct or transitive dependency; 51 vulnerabilities without a fix difficulty are omitted from (a).}
\label{fig:difficulty}
\end{figure}

Transitive vulnerabilities make up 70\% of fixable instances, yet for every harness, passing upgrades fix a smaller share of them than of direct ones (Figure~\ref{fig:difficulty}). Snyk's guidance matters most here: it more than doubles the share of transitive instances Claude Code fixes, from 28\% to 59\%, against 73\% for \remy{}; on direct instances, Claude Code fixes 62\% without it.

\subsection{New Vulnerabilities}
\label{sec:newvulns}

Upgrades can pull in packages with their own known vulnerabilities, which PASS does not penalise (Section~\ref{sec:future}). Post-fix scans show new instances in 62 \remy{} attempts (across 16 repositories), 30 \aided{} and 33 \unaided{} attempts.

\section{Discussion}
\label{sec:discussion}

\paragraph{Dependency upgrades and vulnerability fixes must be evaluated jointly.} Scored on dependency resolution, build and tests alone, the agents perform similarly, because leaving a vulnerability in place rarely breaks anything; scored on the vulnerability alone, \remy{}'s breakages would go unnoticed. A breaking-update benchmark, which checks no vulnerability, therefore cannot observe avoidance, Claude Code's most common failure.

\paragraph{An SCA scanner is a key tool for agents fixing vulnerabilities.} Breaking-update benchmarks already show that agents can partly repair what an upgrade breaks, by adapting code and migrating APIs (Section~\ref{sec:related}); our results show that agents also need vulnerability data. Given the Snyk scan and recommended upgrade, Claude Code more than halves the attempts whose first failed gate is the rescan (107 to 48). Without them, it regressed on 20 upgrades that Snyk's recommended upgrade alone solves, mostly by choosing a version that is still vulnerable, and its passing upgrades fixed less than half as large a share of transitive as of direct vulnerabilities. Web search does not reliably tell an agent which version is safe, least of all for packages the project does not declare directly.

\paragraph{Knowing the target is not enough.} With the same findings and target, \remy{} passes 39 more upgrades than \aided{}, 23 of them major, and rarely shows \aided{}'s main failures: stopping within the current major version and pinning transitive packages instead (Section~\ref{sec:failures}). \remy{} combines Snyk intelligence with a loop in which it builds, tests and repairs (Section~\ref{sec:agents}), and our setup cannot isolate which component makes the difference. That the system around the model can matter this much is consistent with DepRepair, where the full pipeline (distilled migration rules plus localisation and per-category guidance) raised pass rates by 7--11 points while raw upstream evidence lowered them by 7--23 points~\cite{yang2026deprepair}, and with DepBench, where changing only the harness moved results by up to 16.3 points~\cite{luo2026depbench}.

\paragraph{Failure costs and remaining headroom.} Claude Code's failures are mostly silent: the vulnerability was left unfixed, and in 70 of its 124 failures, \unaided{} reported it fixed. Most of \remy{}'s failures are visible in the project's own build and tests, where CI or a reviewer will see them. All agents still struggle with major framework migrations: 23 upgrades, mostly Spring, Spring Boot and Dropwizard migrations, defeated all three, and half of \remy{}'s major PASSes needed application-source edits. \bench{} therefore leaves room for better agents, and major migrations may still warrant human review.

\section{Threats to Validity}
\label{sec:limitations}

\paragraph{Construct validity.} A PASS is only as reliable as the SCA scanner and the project's tests.
\begin{itemize}
  \item \textbf{Snyk Open Source as task source and judge.} Tasks and the rescan use Snyk's vulnerability database, which \unaided{} cannot consult, so a fixed version taken from another database may still be flagged. This penalises only \unaided{}; \remy{} and \aided{} receive the same findings and target.
  \item \textbf{Test strength.} Breaking changes, especially changes in behaviour, are caught only as well as the project's tests exercise them. Six repositories (91 of the 307 upgrades) have at most four passing baseline tests, leaving most of the work to the build gate. \bench{} also relies only on each project's own tests, which agents can run, with no held-out tests or tests generated to target the upgrade. Held-out tests catch failures that visible ones miss: in DepBench, 61.8\% of failed attempts had passed the visible tests or build~\cite{luo2026depbench}. Even purpose-built tests catch only part: BreakGuard, which generates tests specifically to detect breaking dependency updates, catches at best 30.3\% of those in the BUMP benchmark~\cite{raj2026breakguard}.
  \item \textbf{Verifier noise.} Test identities omit the module name, so same-named tests in different modules can mask regressions; three npm repositories without lockfiles let dependency trees float between baseline and verification; and the baseline is measured once, so a flaky test can be recorded as a regression (BUMP reruns tests three times~\cite{reyes2024bump}).
  \item \textbf{A deliberately simple baseline.} The \pmbump{} never moves a BOM or parent or tries another version, so the Snyk-advice-solvable split is a lower bound on what mechanical upgrading can solve, and the agents' gains overstate what only an agent can do (Section~\ref{sec:failures}).
  \item \textbf{New vulnerabilities not penalised.} They appear in more \remy{} attempts than Claude Code attempts (Section~\ref{sec:newvulns}), so this favours \remy{}.
\end{itemize}

\paragraph{Internal validity.}
\begin{itemize}
  \item \textbf{Leakage.} The GitHub and Snyk blocks are host-file based, without egress filtering, so DNS-over-HTTPS, hard-coded IPs, unlisted mirrors, web archives and search snippets can bypass them; more simply, about 25 repositories are published libraries whose later releases are on package registries. Leakage would favour Claude Code, since \remy{} has no web tool and its prompt forbids such fetches; Claude Code transcripts were not kept, so leakage cannot be audited.
  \item \textbf{Training-data exposure.} 40 of the 43 public repositories are pinned at commits that predate Claude Sonnet 5.5's training data cutoff (June 2026)~\cite{anthropicmodels}, so the model may have seen these projects, their later upgrades or fixes for the same vulnerabilities. This may inflate absolute pass rates, but all three agents share the model.
\end{itemize}

\paragraph{External validity.}
\begin{itemize}
  \item \textbf{Case set selection.} A few very large repositories, such as Apache Kafka and Spring Framework, were excluded because they exhausted memory or time in earlier runs or had hundreds of upgrades that would take days to evaluate one by one, so \bench{} under-represents very large repositories and, since several are Gradle builds, Gradle.
  \item \textbf{Scope.} \bench{} covers the six package managers behind about 92\% of production issue--project pairs (not Go, .NET, Ruby, PHP or pnpm), and scores upgrades one at a time rather than ordered and combined in one tree, as remediation is usually run in practice.
\end{itemize}

\paragraph{Conclusion validity.} Each agent, and the \pmbump{}, ran once per upgrade, so the significance tests capture variation across repositories, not reruns. Upgrades are concentrated: the three largest repositories contribute 27\% of them, and single repositories dominate some failure categories (e.g.\ one accounts for 28 of the 30 Claude Code dependency-gate failures), so the pooled counts of Section~\ref{sec:failures} lean on a few projects, whereas macro rates weight repositories equally. All counts are a snapshot of the vulnerability database at scan time (Section~\ref{sec:data}).

\paragraph{Future work.}\label{sec:future} The planned re-run of \bench{} adds repeated runs to measure run-to-run variance; three baseline test runs to detect flakiness; held-out tests that agents cannot run, to catch changes in behaviour; a whole-repository mode that applies all of a repository's upgrades together instead of each in isolation; a verdict that also penalises new vulnerabilities; egress-level leakage controls; further ecosystems such as Go, .NET and pnpm; more models, to separate the effect of the model from that of the agent; and retained transcripts, for failure-cause adjudication by several annotators with measured agreement.

\section{Data Availability}
\label{sec:data}

The \bench{} harness (task builder, sandbox, verifier and analysis scripts) and the 43 evaluation cases (repository, pinned commit, image and commands) are open source at \url{https://github.com/snyk-labs/snyk-vulnbump}. Tasks are derived from a Snyk scan at run time, which requires the Snyk CLI and a token. The repositories are pinned but the vulnerability database is not, so a later run on the same commits will usually find more than our 2{,}066 fixable instances (scanned 4--5 October 2026) and may recommend different targets. For example, CVE-2026-76183 (high severity), published between our last agent scan (5 October) and the \pmbump{} run (8 October), is fixed only from \texttt{tomcat-embed-websocket} 10.1.60, so two upgrades in tokuhirom/java-handbook target 10.1.60 for the \pmbump{} but 10.1.59 for the agents. Runs scanned on different days should therefore be compared only on tasks with the same recommended upgrade, and against their own baseline.

\section{Conclusion}

\bench{} evaluates vulnerability-driven dependency upgrades in 43 real repositories across six package managers, matched to Snyk Open Source production data, with a joint verdict checked independently in a fresh container: the vulnerability must be fixed and the project must still resolve, build and pass its tests. On 307 upgrades with the same model, this verdict separates agents that a breakage-only check ranks equal: \remy{} passes 90.2\% of upgrades, against 77.5\% for \aided{} and 59.6\% for \unaided{}. Claude Code more often leaves the vulnerability in place while \remy{} more often lands the upgrade, occasionally breaking the project. An SCA scanner is a key tool for remediating vulnerable dependencies: Snyk's guidance raises Claude Code's PASS rate by 17.9 percentage points, and without it Claude Code often settles on a version that is still vulnerable. Yet knowing the target is not enough: given the same Snyk findings and model, \remy{}'s purpose-built harness passes 12.7 points more than \aided{}. Major framework migrations remain hard for every agent. Repeated runs, held-out tests and a whole-repository mode are the next steps towards a more reliable and realistic evaluation of vulnerability-driven dependency upgrades.

\bibliographystyle{plain}
\renewcommand{\bibfont}{\raggedright}
\bibliography{vulnbump}

\end{document}